\documentclass{article}
\usepackage{arxiv}
\usepackage{colortbl}
\usepackage{comment}
\usepackage{authblk}
\makeatletter
\newcommand*{\rom}[1]{\expandafter\@slowromancap\romannumeral #1@}
\makeatother
\usepackage{cite}
\usepackage[pdftex]{graphicx}
\usepackage{url}
\usepackage{color}
\usepackage{graphicx}
\usepackage{hyperref}

\usepackage{multirow}
\usepackage{amsmath}
\usepackage{algorithmic}
\usepackage[ruled,norelsize]{algorithm2e} 

\SetAlFnt{\small}
\SetAlCapFnt{\small}
\SetAlCapNameFnt{\small}
\SetAlCapHSkip{0pt}
\IncMargin{-\parindent}

\usepackage{mathtools}

\allowdisplaybreaks

\begin{document}

\title{\Large SIMULATION-BASED DIGITAL TWIN DEVELOPMENT FOR BLOCKCHAIN ENABLED END-TO-END INDUSTRIAL HEMP SUPPLY CHAIN RISK MANAGEMENT}
\author[1]{Keqi Wang}
\author[ ,1]{Wei Xie\thanks{Corresponding author: w.xie@northeastern.edu}
}
\author[2]{Wencen Wu}
\author[1]{Bo Wang}
\author[1]{Jinxiang Pei}
\author[3]{Mike Baker}
\author[4]{Qi Zhou}
\affil[1]{Northeastern University, Boston, MA 02115}
\affil[2]{San Jose State University, San Jose, CA 95192}
\affil[3]{Willamette Valley Assured LLC., Lebanon, OR 97355}
\affil[4]{QuarkChain Inc., San Mateo, CA 94401}

\maketitle

\begin{abstract}
With the passage of the 2018 U.S. Farm Bill, Industrial Hemp production is moved from limited pilot programs to a regulated agriculture production system. However, Industrial Hemp Supply Chain (IHSC) faces critical challenges, including: high complexity and variability, very limited production knowledge, lack of data and information tracking. In this paper, we propose blockchain-enabled IHSC and develop a preliminary simulation-based digital twin for this distributed cyber-physical system (CPS) to support the process learning and risk management. Basically, we develop a two-layer blockchain with proof of authority smart contract, which can track the data and key information, improve the supply chain transparency, and leverage local authorities and state regulators to ensure the quality control verification. Then, we introduce a stochastic simulation-based digital twin for IHSC risk management, which can characterize the process spatial-temporal causal interdependencies and dynamic evolution to guide risk control and decision making. Our empirical study demonstrates the promising performance of proposed platform.
\end{abstract}


\section{INTRODUCTION}

\textit{The \underline{objective} of our study is to develop a simulation-based cyber-physical system (CPS) digital twin for blockchain-enabled industrial hemp supply chain (IHSC), which will be utilized to improve the understanding of end-to-end process, guide quality control verification, and accelerate the development of a safe, efficient, reliable and automated supply chain system.} 
With the passage of the 2018 U.S. Farm Bill, IH production is moved from being limited pilot programs administered by state regulatory officials 
and becomes a public \textit{regulated} agriculture production. Due to the huge success of farmers who participate in the pilot programs in various states, there is intense interest in growing IH, and the total area licensed for hemp production has increased dramatically from 37,122 (in year 2017) to 310,721 (in year 2019) acres nationally (from 3,500 to 51,313 acres in the state of Oregon) 
\cite{sterns2019emerging}. 
According to the U.S. Department of Agriculture (USDA), the sales are expected to increase from \$25 million in 2020 to more than \$100 million by 2022. 
However, there was very limited knowledge/experiences and well-developed guidelines 
on how to navigate the end-to-end IHSC process, including seed selection, fertility, harvesting and handling on an industrial scale.

The two most common products produced from hemp are cannabidiol (CBD for medicinal uses) and tetrahydrocannabinol (THC for recreational drug use).
As opposed to the safety of CBD, THC has been reported with unpleasant side-effects including but not limited to anxiety and panic, impaired attention and memory 
\cite{hall1998adverse}.
Considering the major active ingredient in marijuana causing psychoactive effects, the law (2018 U.S. Farm Bill) requires that CBD plants and derivatives should contain no more than 0.3 percent THC on a dry weight basis. 
{Thus, the development of a safe, reliable, sustainable, efficient and automated IHSC plays a critical role in improving this economy and ensuring public health.}

\textit{The management of IHSC as a cyber-physical system faces critical \underline{challenges}, including: high complexity and variability, very limited IHSC knowledge, data tampering, scalability, data and information tracking.} 
First, the IH supply chain process is complex and dynamic. The content flow of CBD/THC  
highly depends on the complex interactions of many physical/human factors in seed selection, cultivation, stabilization, and processing phases, such as seed varieties, soil properties, weather condition, drying and extraction. 
Second, the data collection cycle 
is lengthy (about one year), 
and the existing IH production knowledge is very limited. 
Third, the data integrity is hardly guaranteed without human intervention. Fourth, how to allocate the limited inspection resources and monitor this fast growing industry is extremely challenging. 
\textit{Therefore, IH practitioners and government regulators have urgent \underline{needs} in improving supply chain transparency, developing comprehensive/deep understanding of the end-to-end IHSC, eliminating risks, ensuring CBD product quality, and controlling the THC content flow in the supply chain.}

Driven by these challenges and needs, we propose blockchain-enabled IHSC and create a digital twin of this CPS to support the process surveillance, learning, and risk management. 
We explore the two-layer blockchain with proof of authority (PoA) and smart contract consensus \cite{xie2019simulation}, which can efficiently leverage local authorities (e.g., Willamette Valley Assured LLC.) and limited officials inspection resources (e.g., State Regulatory) 
to improve the supply chain transparency and ensure the data integrity. 
We further develop a stochastic simulation for the end-to-end blockchain-enabled IHSC from seed selection to the final commercial product (CBD oil). We consider the complex interactions of random input factors and decisions introduced in different steps and model the dynamic evolution of key attributes of IH product (e.g., CBD and THC). It can characterize the spatial-temporal process interdependencies and provide the reliable guidance on risk analysis/control and decision making for IHSC. 

\textit{This study is based on our academia-industry collaboration on ``AI- and Blockchain-based IoT Platform Development for End-to-End IHSC Learning, Risk Management, and Automation."} 
Our blockchain is built on QuackChain network (see \href{https://github.com/QuarkChain/pyquarkchain/wiki/Cluster-Design}{quarkchain wiki}). We develop a user-friendly mobile app so that each participant can use smart phone to real-time collect and upload their data to the cloud, and further share the process verification and tracking information through the blockchain network. 
The recently finished blockchain-enabled IHSC platform and digital twin are being tested and validated during the small-scale pilot phase: IH season 2020 (May 15 - November 30) in Oregon and Pennsylvania. 
The collected process data will be studied to improve the IHSC knowledge and decision making, guide the knowledge-based reputation learning, improve the digital twin development, and accelerate the end-to-end IHSC automation. 

This paper is organized as follows. In Section~\ref{sec:IHSC}, we provide the problem description and introduce the end-to-end blockchain-enabled IHSC. In Section~\ref{sec:blockchain}, we discuss the two-layer blockchain and PoA smart contract design for process data tracking and information sharing. In Section~\ref{sec:simulationModeling}, we create the simulation model for blockchain-enabled IHSC and provide the process risk analysis. After that, we conduct the empirical study in Section~\ref{sec:empiricalStudy} and conclude this paper in Section~\ref{sec:conclusion}.

\vspace{-0.1in}

\section{END-TO-END BLOCKCHAIN-ENABLED INDUSTRIAL HEMP SUPPLY CHAIN}
\label{sec:IHSC}
\begin{figure}[htb]
{
\centering
\includegraphics[width=1\textwidth]{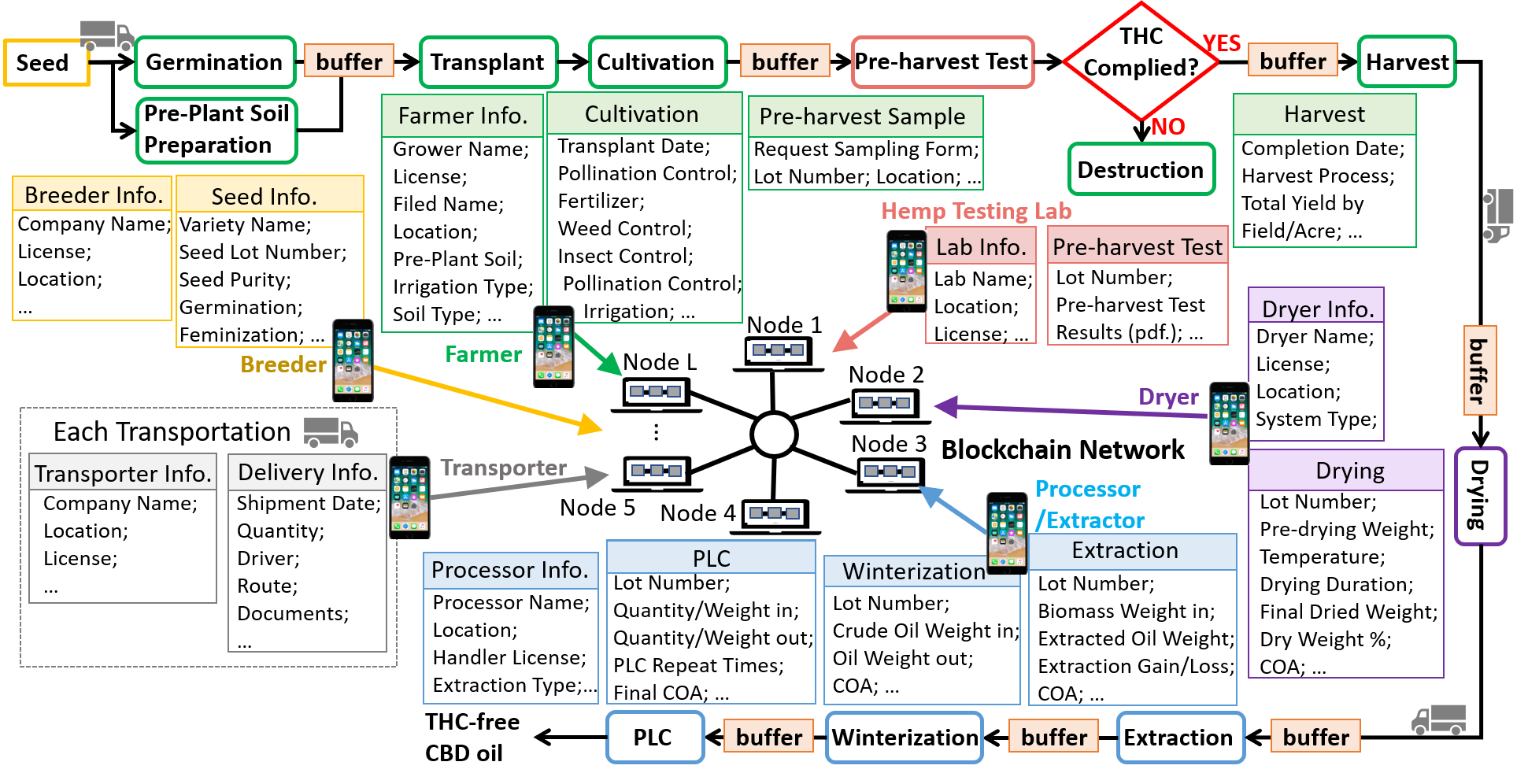}
\vspace{-0.3in}
\caption{Blockchain-Enable End-to-End Industrial Hemp Supply Chain \label{fig: IHSC}}
}
\end{figure}

Even though the proposed simulation modeling and blockchain-based CPS platform are applicable to general regulated manufacturing and supply chain, in this paper, we focus on the IHSC from seed selection to final commercial product, called ``THC free broad-spectrum CBD oil", which will be shortened as \underline{CBD oil} for simplification.  The main operation units of IHSC include: (1) cultivation, (2) stabilization (drying), and (3) manufacturing (extraction, winterization, and purification-PLC); see Figure~\ref{fig: IHSC}. Since farmers have strong incentive to maintain the lot uniformity, the same seed variety and growing strategy are typically applied to the same lot.
Thus, we ignore the IH plant variation in the same lot,  
and consider the  lot-to-lot variation of IHSC outputs, including productivity, quality, and cycle time.

\noindent \underline{\textbf{Cultivation:}} It starts with selection of seed variety, 
and licensed growers (farmers) select one or a few IH seed varieties to grow for the incoming season. Here, we consider one particular variety: Special Sauce provided by \href{https://oregoncbdseeds.com/resources/}{Oregon CBD}.
Once receiving the selected seed, the farmer simultaneously starts Germination and Pre-Plant Soil Preparation process. 
Germination happens at the greenhouse with the controlled-growth environment,
where the seed germinates (the leaves shoot up and the roots dig down) in about 5--10 days. After germination, there is at most 2-day holding period before transplanting. 
Otherwise, it is abandoned.  It is important that the seedlings are not stressed after leaving the greenhouse and being planted. At the meantime, after the Soil Test (which provides a detailed and comprehensive description of soil, such as pH, nutrition level, heavy metal, herbicide and pesticide contains), the farmer performs Pre-Plant Soil Preparation to adjust the properties of field soil, and this process usually takes 1--2 days. Once both steps are finished, the germinated seedlings are transplanted into the field, which takes about 1--2 days. 

Then, the farmer continues the cultivation process including two typical sub-phases: Vegetative and Flowering. The Vegetative is the main growth stage, where the iconic fan leaves begin to develop and grow larger with more blades and fuller sets of leaves. The Flowering is the final stage of growth, where the resinous buds develop (the majority of the CBD is generated). Once the flowers begin to develop, they require a different kind of care. 
Besides the regular control of weed, insect and mold, the most important one is pollination control. If the plants are pollinated, they shift their energy from producing unpollinated flowers that are high in oil to pollinated flowers that consume oil to produce seed, which will decrease the CBD yield significantly. Overall, the cultivation process in total takes about 50--60 days. 

After that, the sample of IH is sent 
to Hemp Testing Lab for \textit{pre-harvest test} that takes 2--7 days on average and gives the comprehensive test results including CBD/THC levels. \textit{Based on the USDA regulation, the lot with THC level greater than 0.3\% is required to be destroyed completely.} For IH lots that have passed the pre-harvest test, farmers can start the harvest process. Usually farmers tend to wait for several days since the CBD level further increases during this period. However, since the THC content also increases, the government requires farmers to complete harvest no longer than 15 days after the pre-harvest test. 
If farmer doesn't finish harvesting these lots in time, the rest needs to be scheduled for another testing and then followed by new harvest within another 15-day limit. Thus, farmers need to make an appropriate harvest schedule to consider risk-profit trade-off.

\vspace{0.03in}
\noindent \underline{\textbf{Stabilization:}} 
The harvested IH biomass often has a high moisture content, even up to 85\%, which has to be reduced to an ideal level to avoid deterioration in 
storage and transportation. Typically, there are two types of stabilization: drying-stabilizing and freezing-stabilizing. Since drying is most common for the post-harvest stabilization process, we consider it here. Based on the drying 
machine availability, the harvested IH biomass may have to wait. The probability of being contaminated increases during waiting with high moisture, which can cause lost in profit.
The drying process typically takes 1-2 days.

\vspace{0.03in}
\noindent \underline{\textbf{Manufacturing:}}
The dried IH biomass is then sold to licensed processors/extractors to produce the desired commercial product (CBD oil). The main steps include Extraction, Winterization, and Preparatory Liquid Chromatography (PLC). Since the dried biomass has life of 9--12 month, the product can wait in buffer before each step. Extraction is a process to produce crude oil, which typically includes two steps: extraction and decarboxylation. 
The Ethanol Extraction is one of the most efficient  methods for processing large batches of IH biomass. This method cycles ethanol solvent through the solid biomass, strips the cannabinoids, terpenes and plant waxes from the flower to produce the CBD crude oil. 
Although CBD is the main cannabinoids with the healthy and medicinal potential, 
the highest concentrations of cannabinoid in biomass (flowers) is cannabidiolic aid (CBDA). After extraction process, the decarboxylation reaction is needed to transform CBDA into its neutral cannabinoids CBD  \cite{wang2016decarboxylation}. Also, the decarboxylation can prevent the degradation of desirable cannabinoids \cite{romano2013cannabis}. Thus, the oil is to be processed under controlled temperature to decarboxylate the CBDA to CBD. In the mean time, the tetrahydrocannabinolic acid (THCA) is transformed as THC. Not only the cannabinoids but also other undesirable element in biomass are extracted. The whole Extraction process usually takes 1--2 days.

Then, another process, called winterization, needs to be applied to remove the undesirable fat, terpene and wax from crude oil. During this process, the content of the CBD/THC stays almost the same with very minor drop in absolute weight, and the percentage of CBD/THC increases. This process takes about 1--3 days. Finally, several PLC steps are used to remove some specific target molecule, such as THC, and remediate the oil with THC level below the standardized requirement. Considering the economically feasibility, we assume at most twice of this purity process in the empirical study. After the PLC steps, there will be another Certificate of Analysis (COA) test, and the final product can only be acceptable at less than 0.05\% THC (for THC free broad-spectrum CBD oil).

\vspace{-0.1in}

\section{BLOCKCHAIN BASED DATA/INFORMATION TRACKING SYSTEM}
\label{sec:blockchain}

Here we present the two-layer blockchain based data tracking, information sharing, and interoperability framework for the end-to-end IHSC (see Figure~\ref{fig: IHSC}), which can greatly improve both security and efficiency. In Section~\ref{sec:participants}, we first describe 
the important data/information collected by each group of participants. Here, we consider a static setting with fixed participants.
In Section~\ref{subsec:blockchain},
we review the two-layer blockchain design and the PoA based smart contract proposed in our previous study \cite{xie2019simulation}, and then present how to apply it to the end-to-end IHSC to support interoperability and improve safety and efficeincy.

\vspace{-0.1in}

\subsection{Participants of IHSC and Data Collection}\label{sec:participants}

Here, we list main participants of end-to-end IHSC.
By tracking all the critical information/data provided by each participant, we are able to keep the comprehensive monitoring over the process. 

\vspace{-0.0in}
\noindent \underline{\textbf{Breeder:}} 
A breeder produces one or several IH seed varieties or clones, which will be sold to the farmers. Prior to planting and breeding, breeders are required to register and provide the information/data as follows. 

\vspace{-0.12in}
\begin{itemize}
    \item \textit{Breeder registration information}: breeder name, registration, certificate, location and so on. This information provides a unique reference to each breeder.
    
    \item \textit{Seed type/source}: variety name and description, seed lot number, seed purity analysis, pure seed percentage, seed germination percentage, seed germination data completed, flowering type (photoperiod or autoflower), ferminization process, seed ferminization percentage, clone information. The seed type/source data will be included in each transaction data when a farmer purchases any seed/clone variety from a breeder.
    
\end{itemize}

\vspace{-0.1in}
\noindent \underline{\textbf{Licensed Grower:}} 
A grower obtains the seeds varieties from breeder, and then cultivate the IH until harvest. Before growing IH, farmers are required to have certificate and provide the information as follows. 
\vspace{-0.15in}
\begin{itemize}
    \item \textit{Grower information}: The name and license information provide a unique reference to each grower.
    
    \item \textit{Field information}: field name, GPS field location, approved background check, pre-plant soil tests, irrigation type, prior year field history, soil type. 
    
    \item \textit{Cultivation data}: seeding/transplanting data, plant density, row width, grown on plastic, irrigation type/frequency/volume, fertilizer frequency/volume, weed control, insect control, mold control, pollination control. During the germination, transplant and cultivation process, the critical cultivation data, i.e., field/IH plant pictures/image data will be taken timely and uploaded to keep a record of the growing process for each lot. 
    
    \item \textit{Pre-harvest test request}: 
  Pre-harvest Hemp Sampling and Testing Request Form. The growers fulfill the document and send it to the approved lab requesting a lot to be sampled. Typically, this form includes lot number, GPS location and acres.
    
    \item \textit{Harvest data}: harvest process and completion dates, moisture content, total yield by field. 
    After receiving the pre-harvest test results, following the 0.3\% THC commitment and instructions, the grower will either destroy or start to harvest the corresponding IH lots and upload the harvest information to blockchain. \textit{At this stage, the data provided by growers has to be \underline{verified and confirmed} by local authorities and regulatory officials.} 
    
\end{itemize}

\vspace{-0.1in}
\noindent \underline{\textbf{Dryer:}} 
A dryer provides the harvested IH drying-stabilization service for the growers. The dryers need to provide the drying-stabilization data and also post stabilization test results.
\vspace{-0.12in}
\begin{itemize}
    \item \textit{Drying-stabilization data}: dryer name/location/license, type of drying system, lot number, 
    pre-drying weight, temperature/duration of drying, final dried weight, dry weight, container type/weight, brushing to remove excess stalks. It provides the drying process record for each lot of IH.
    
    \item \textit{Post stabilization test}: cannabinoid content (CBD and THC), pesticide residue, heavy metals. The COA test is needed for trading the biomass to an extractor or the crop analysis records for vertically integrated organizations. At this stage, THC can be higher than 0.3\% 
    and the biomass must be tracked. 
    
\end{itemize}

\noindent \underline{\textbf{Licensed Processor/Extractor:}} 
A processor purchases the dried IH biomass from the growers/dryers, and then conduct the processing steps to produce commercial CBD oil. He needs to provide the extraction data, winterization data, PLC data, and the post PLC COA test results. \textit{After the post PLC COA test, the data/information provided by the processors needs to be \underline{verified and confirmed} by local authorities and regulatory officials.} The verified data will be included in new blocks and attached to the blockchain.

\begin{itemize}
    \item \textit{Extraction data}: processor-extractor name/location, handler license information, extraction type, lot number, biomass weight in, extraction input quantity/recaptured, quantity of oil extracted, gain/loss of extraction, post extraction test. The extraction data provides a record of processor/handler information (i.e., responsible operator) and extraction process information.
    
    \item \textit{Winterization data}: lot number, crude oil weight in, winterized oil out, and post winterization test. 
    The processors need to provide the record of the winterization process on the crude oil.
    
    \item \textit{PLC data}: lot number, quantity/weight in, quantity/weight out, PLC repeat times, and post PLC test. After the PLC COA test, all the data collected from the processor need to be verified by authorities and further uploaded to the blockchain. The finished commercial CBD oil, with the comprehensive and reliable record of the end-to-end IHSC, can enter the consumer market. 
    
\end{itemize}

\noindent \underline{\textbf{Transporter:}} 
A transporter delivers the items (intermediate IH product) between different participants of the IHSC. He needs to provide the transportation data, including sender, receiver, and delivery information. 

\begin{itemize}
    \item \textit{Transportation data}: transporter information, driver license, lot number, sender/receiver information, quantity, shipment/delivery date, vehicle/model of transport, route of transportation. By tracking the transportation data, the system keeps the record of comprehensive delivery information for each product/IH biomass. In addition, drivers will use the corresponding delivery information as the legal permission for carrying IH product under road investigation.
    
\end{itemize}

\noindent \underline{\textbf{Hemp Testing Lab:}} 
A Hemp Testing Lab receives the pre-harvest IH samples from the official sampler and then provide the pre-harvest test results and send back to the growers. 

\begin{itemize}
    \item \textit{Lab information}: Lab name, location, and other related information, which will provide a unique reference to each lab.

    \item \textit{Pre-harvest test}: sampling date, lot number, COA test, 
    cannabinoid content (CBD and THC), pesticide residue, heavy metals, etc. The official sampler will take the sample from the field after the Lab receiving the request form. He completes the Pre-Harvest Hemp On-site Sampling form and sends the sample to the Lab for analysis. 
    Then the Hemp Testing Lab will upload the pre-harvest results for each lot sample to blockchain, and send the copy back to the grower. 
    
\end{itemize}

\vspace{-0.1in}

\noindent \underline{\textbf{Authorities/Regulators:}} 
The local authorities are responsible for the verification in their local areas, 
and then the regulators (i.e., state USDA) will confirm all the uploaded data (may randomly select some for detailed verification). Regulatory officials will 
upload the confirmed data and decision to the blockchain. \textit{Suppose the onsite verification takes much more time than online confirmation.}

\begin{itemize}
    \item \textit{Authorities information}: the company information, location, certificates, and other information, which will provide a unique reference to each group of local authorities.
    
    \item \textit{Verification/confirmation results}: the local authorities and regulatory officials will provide the verification and confirmation decision, as well as the date/time, and responsible persons or group.  
    
\end{itemize}


\subsection{Two-Layer Blockchain to Support IHSC Interoperability and Safety}
\label{subsec:blockchain}

\begin{figure}[htb]
{
\centering
\includegraphics[width=0.7\textwidth]{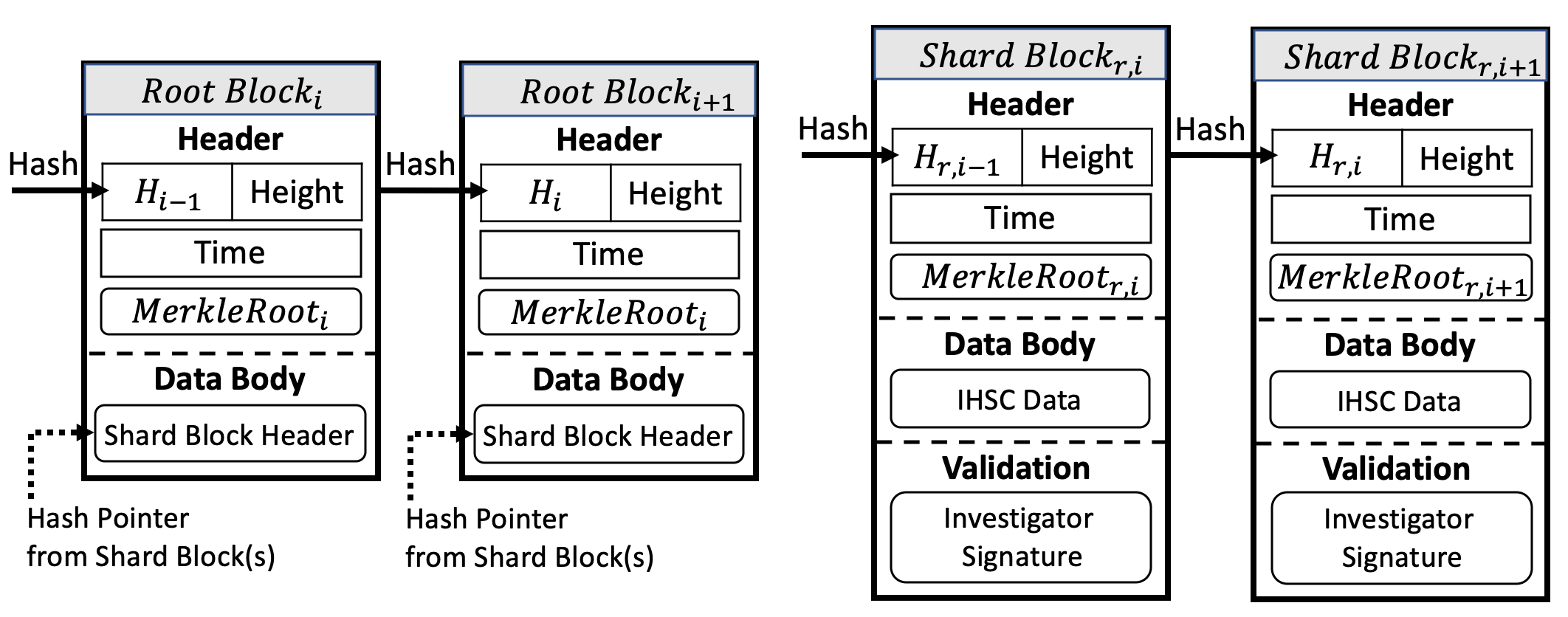}
\vspace{-0.15in}
\caption{The root (on the left) and shard (on the right) chains structure. \label{fig:shardblock}}
}
\end{figure}

Here, we briefly review the two-layer blockchain and the PoA based smart contract proposed in our previous study \cite{xie2019simulation}, and then discuss how to extend it to integrated IHSC. The blockchain is a distributed database, also a global ledger that records all the process data in the network as a timestamp chain of blocks. \textit{Each block contains three parts: header, data body, and validation;} see Figure~\ref{fig:shardblock}. More specific, the header records the block created time, the height of the current block, the ``Nonce" number used to encipher the hashed block, and also the ``Merkle Root" that saves the hash of several data/documents (e.g., image or pdf files) and validation information. Each block references to the hash of the block that comes before it, which establishes a chain of the blocks (called blockchain). When new valid data/information records need to be written in the global ledger, new blocks will be generated and added to the blockchain, and broadcast to the network so that each node will update the copy of the ledger. Therefore, the blockchain technology provides the data tracking and information sharing, which can fulfill the needs for IHSC, including transparency, consistency, interoperability, safety, and data integrity.

To facilitate the automation and support the participants reaching a common global view of the world state, each blockchain network needs to establish consensus rules that each data record (or transaction) should conform to. It can be achieved by smart contract \cite{bahga2016blockchain}. 
Any participant can send new data/information record to the contract for the verification, 
and then the verified data/information is broadcast to each node in the blockchain network to reach global consensus. For the highly regulated IHSC, Proof-of-Authority (PoA) smart contract is used to ensure that the blockchain contains valid and reliable data/information and reduce the impact of data tampering and cyberattacks. Specifically, each record will be assigned (by smart contract) to an authority node (e.g., state regulators) for verification, and then the verified data are uploaded to blockchain. In our project, \underline{on-site investigations} are required to ensure that the data records of IHSC (i.e., images and reports) are accurate and reliable.

However, for typical one-layer blockchain structure and current IHSC design with only state regulators in charge of inspection, the verification efficiency totally depends on the number of available officials for on-site visit. 
Given the limited inspection resources at state USDA, it is challenging to handle the situation when the scale of the IH market grows
dramatically.
\textit{Therefore, we consider the two-layer blockchain platform which can leverage the additional resources from local authorities with state regulators, and improve the efficiency and security of IHSC; see the structure illustration in Figure~\ref{fig:shardblock}.} The two-layer blockchain includes: (1) the sharding layer supported by local authorities for verification, which divides the IHSC or IH market into multiple area/location-based shards; and (2) root chain layer supported by state regulators, which serves as the coordinator among all shard chains. 
We use the root-chain-first PoA which has a \textit{hierarchical verification procedure}. When any process data/information records (e.g., pre-harvest test results, harvest data) need to be uploaded to the blockchain, the smart contract on corresponding shard will send verification request to the local authority, that will then dispatch investigator for on-site validation. The validated data together with investigator's signature will be included in new shard block, which will be attached to the shard chain and broadcast to the network. After that, the state regulator will review and online confirm the new uploaded data/information, which is relatively quick.
Once complete confirmation, a new root block including the corresponding shard block header will be generated and append to the root chain, and the data/information is viewed as ``authentic" in the IHSC blockchain network. \textit{The state sharding technology and root-chain-first consensus enable different local authorities to simultaneously process  the data coming into different shards, which can improve the IHSC efficiency and interoperability.}

\vspace{-0.08in}

\vspace{-0.2in}

\section{SIMULATION-BASED DIGITAL TWIN FOR BLOCKCHAIN-ENABLED IHSC}
\label{sec:simulationModeling}

We develop the simulation model as digital twin of IHSC cyber-physical system, which can characterize the interactions of random input factors and quantify the output variability.
It can facilitate the risk management of blockchain-enabled IHSC and guide dynamic decision making. We can track the dynamic evolution of key attributes (i.e., CBD and THC) during the end-to-end IHSC. Since the IH is seasonal plant, we assume all the lots 
receive the seeds at the same time for each growing season. Thus, we consider the deterministic arrivals of certain lots 
(say $n$) of seeds, with inter-arrival time equals to the grow cycle (i.e., 365 days). 

\vspace{-0.1in}
\subsection{Input Modeling}
\label{subsec:inputModeling}

Given limited knowledge and historical data of IHSC process, we develop the input models characterizing the main sources of uncertainty in physical and cyber parts of blockchain-enabled IHSC. 

\vspace{-0.1in}
\subsubsection{Cultivation Process}

After receive the seeds, the farmers will start Germination and Pre-Plant Soil Preparation simultaneously. 

\vspace{0.0in}
\noindent {\textbf{Germination:}} 
During germination, there's no CBD and THC, we model the germination time following uniform distribution $\mbox{Unif}(5, 10)$ days. Then, the germinated IH seedlings need to wait in buffers until the field is available. If the seedling wait time is more than 2 days, this lot will be dropped.

\vspace{0.0in}
\noindent {\textbf{Pre-Plant Soil Preparation:}} 
For pre-plant soil preparation, suppose farmers have already done the soil test and known their soil properties. 
We model the preparation time following uniform distribution $\mbox{Unif}(1, 2)$.

\vspace{0.0in}
\noindent {\textbf{Transplant:}} The germinated IH seedlings will be transplanted into the field. We model the transplant time following uniform distribution $\mbox{Unif}(1, 2)$.

\vspace{0.0in}
\noindent {\textbf{Cultivation (Vegetative and Flowering):}}
Due to lack of historical process data, we model Vegetative and Flowering together as the Cultivation process following uniform distribution $\mbox{Unif}(50, 60)$. This is the main process that develops the CBD and THC, and the growth processes are modeled by the linear models,
\begin{equation}\label{eq.growth}
U_c = (gt + \epsilon)\dfrac{r}{r+1}, ~~
V_c = (gt + \epsilon)\dfrac{1}{r+1}
\end{equation}
where $U_c$ and $V_c$ denote the CBD and THC levels after cultivation step respectively, $g$ is the growth/cumulative rates of total cannabinoid content (CBD + THC) in percentage level (see \href{https://oregoncbdseeds.com/resources/}{Oregon CBD}),  $t$ is the flowing time, $r$ is the ratio of CBD to THC (for the particular seed variety we consider, $r=28:1$). The growth intrinsic uncertainty  $\epsilon \sim {N}(0, gt\lambda^2)$ modeled by truncated Normal distribution quantifies the lot-to-lot variation in total cannabinoid (i.e., CBD and THC) level, which is induced by various factors, including weather, soil, moisture, and seed properties, and farmer’s growing skill.

\vspace{0.0in}
\noindent {\textbf{Pre-Harvest Test:}}
After the flowering, the official sampler (working for lab)
will take samples of each lot and send to Hemp Testing Lab for pre-harvest test. The law requires that uniform samples are taken so that they can  represent 
the true average total THC in the lot. 
 We model the testing time following uniform distribution $\mbox{Unif}(2, 7)$. Based on the equipment availability, the IH samples may wait in the buffer for testing.
 If certain lot has the THC level greater than the required threshold $\gamma_{v} = 0.3\%$, the IH in that lot must be destroyed entirely. No harvest can begin until the pre-harvest test results are received.

\vspace{0.0in}
\noindent {\textbf{Harvest:}}
After receiving the pre-harvest test results (and passed the test), the farmer could start the harvest procedure. \textit{During the period from taking the pre-harvest sample to finish harvest, denoted by $t^{\prime}$, both CBD and THC will continue to increase.} The government requires that the time for completing the harvest can not exceed 15 days, i.e., $t^{\prime} \le 15$. If farmer doesn't finish harvesting these lots in time, the rest needs to be scheduled for another testing and harvest later. 
Therefore, the farmer may need to make an appropriate harvest schedule, balancing the profit from more CBD and the randomness of pre-harvest test and harvest times.
We model the harvest process time following uniform distribution $\mbox{Unif}(1, 2)$. The increments of CBD and THC still follow the same linear growth model in Equations~\eqref{eq.growth}. So we have,
$
U_h = U_c + (gt^{\prime} + \epsilon^{\prime}){r}/(r+1)$ and $
V_h = V_c + (gt^{\prime} + \epsilon^{\prime})/(r+1)
$,
where $U_h$, $V_h$ denotes the CBD and THC level after harvest step respectively, and the lot-to-lot variation in CBD and THC level is modeled by truncated Normal distribution $\epsilon^{\prime} \sim {N}(0, gt^{\prime}\lambda^2)$. The harvested IH biomass will be stored in buffer and wait for stabilization (drying). Notice that due to high moisture, too much time spent on waiting for drying will increase the probability of contamination and degradation. In simulation, if the harvested IH biomass waits for more than 3 days before drying, we directly discard this lot of product.

\vspace{-0.1in}
\subsubsection{Stabilization Process}

For the stabilization process, we mainly consider the drying procedure. The main purpose of drying is to remove the moisture from the harvest IH biomass, and it does not impact on the CBD and THC content. 
We model the drying time for each lot following uniform distribution $\mbox{Unif}(1, 2)$. We denote the number of drying machines by $N_d$, which is a process decision parameter. Notice that the dried IH biomass becomes much stable with a lifetime of 9-12 month. Thus, in the following processes, we can ignore the degradation of cannabinoid and contamination of IH biomass during waiting in the buffers. The farmer will sell the dried IH biomass to the processors/extractors, that will start to produce the CBD oil. 

\vspace{-0.1in}

\subsubsection{Manufacturing Process}

After the processors/extractors obtain dried IH biomass from farmers, they will perform the extraction, winterization, and one or two PLC steps in sequence. Before each step, the intermediate IH product can wait in buffers until the corresponding equipment is available for the next step. 

\vspace{0.0in}
\noindent {\textbf{Extraction (Extraction and Decarboxylation):}}
Suppose decarboxylation begins right after the extraction. When there is extraction vessel 
available, we can start the extraction process to obtain the CBD-riched crude oil. We model the extraction time following uniform distribution $\mbox{Unif}(1, 2)$. The mass extracted (i.e., CDB and THC) is a random proportion of the original amount, 
$
U_e = Q\cdot U_h$ and $
V_e = Q\cdot V_h,
$
where $Q\sim \mbox{Unif}(0.6, 0.8)$. The extracted crude oil will wait in the buffer until the winterization equipment available.

\vspace{0.0in}
\noindent {\textbf{Winterization:}}
Since both cannabinoids and waxes present in biomass need to be extracted.  The winterization is performed to remove undesirable fat, terpene and waxes from crude oil. We model the winterization process time with uniform distribution $\mbox{Unif}(1, 3)$. During this procedure, 
we assume the random proportion of CBD/THC is removed,
$
U_w = W\cdot U_e$ and 
$V_w = W\cdot V_e$,
where $W\sim \mbox{Unif}(0.95, 1)$. The 
CBD oil after winterization will wait in the buffer until the PLC equipment available.

\vspace{0.0in}
\noindent {\textbf{Preparatory Liquid Chromatography (PLC):}}
Since our target product is THC free broad-spectrum CBD oil, one or two PLC steps are applied to remove THC, while trying to keep CBD. We model each of the PLC process time following uniform distribution $\mbox{Unif}(1, 5)$. Each step of PLC removes the random proportion of CBD/THC,
$
U_f = Q_u\cdot U_w$ and 
$V_f = Q_v\cdot V_w$, 
where $Q_u\sim \mbox{Unif}(0.9, 1.0)$, and $Q_v\sim \mbox{Unif}(0.3, 0.5)$. Notice after the PLC steps, a COA test needs to be performed on the finished CBD oil. The final product can only be acceptable if the percentage of THC is less than the requirement
$\gamma = 0.05\%$.

\vspace{-0.1in}
\subsubsection{Tracking and Verification through Blockchain Network}

To develop the digital twin of blockchain-enabled IHSC, we further create the simulation for the blockchain network. Suppose each participant 
corresponds to a
node $N_i$ of the blockchain network; see the illustration in Figure~\ref{fig: IHSC}. When any node (e.g., farmer, processor) sends the data/transaction to the smart contract for verification, the smart contract will assign it to local authorities and wait for verification results. After verification, a new block including the verified data/transaction will be generated and appended to the shard chain. The verified data/transaction then requires further (but quick) confirmation/documentation by a state USDA regulator. After that, a new root block including corresponding shard block header will be generated and attached to the root chain. At current stage, the regulated IHSC requires on-site verification by authorities, which will be partially 
replaced by automatic verification in our future work, through the knowledge model learned from historical data. Compared to the time of verification and confirmation, the delay in network transmission is negligible, which will be ignored for simplification.

\vspace{0.0in}
\noindent {\textbf{Verification:}} For each shard chain having one local authorities group, we model the verification process as a multiple servers queue, with the number of resources (i.e., officers for verification) denoted by $n_s$. We model the  verification time following exponential distribution $\mbox{Exp}(\mu_v)$, with mean $\mu_v = 0.1$ day.

\vspace{0.0in}
\noindent {\textbf{Confirmation:}} Similarly, we also model the confirmation process in root chain as a multiple servers queue, with the number of resources (i.e., state USDA regulators for confirmation) denoted by $n_r$. We model the confirmation time following exponential distribution $\mbox{Exp}(\mu_c)$, with mean $\mu_c = 0.05$ day.

\vspace{-0.1in}

\subsection{End-to-End IHSC Process Uncertainty Quantification and Risk Analysis}
\label{sec:uncertainty}

\begin{sloppypar}

\textit{The computer simulation developed as digital twin of real blockchain-enabled end-to-end IHSC is utilized to conduct the process risk analysis and guide the decisions to improve the system safety and efficiency.}
The simulation output of IHSC depends on the complex interactions of various random inputs. 
Suppose that there are $L$ sources of uncertainty characterized by the input models, $F = \{F_1, \ldots, F_L\}$, where $F_\ell$ with  $\ell=1,2,\ldots,L$ quantifies the uncertainty from physical/cyber parts of blockchain-enabled IHSC. 
Thus, the vector $\pmb{Z} = \{Z_1, \ldots, Z_L\}$, representing the collection of all random inputs, can monitor the IHSC process and impact on the outputs variability, where $Z_{\ell}\sim F_{\ell}$. Denote the decisions by $\pmb x = (x_1, x_2, \ldots, x_K)^\top$, i.e., harvest/drying scheduling and amount of resources allocated at each operation unit. Then, the simulation output can be written as a function of decisions and randoms inputs, denoted by $Y(\pmb{x}; \pmb{Z})$, such as final CBD oil percentage, THC level, and cycle time of each lot. 
The lot-to-lot output variation is characterized by variance $\mbox{Var}[Y(\pmb{x}; \pmb{Z})]$, which can be estimated by the simulation outputs $\{Y(\pmb{x}; \pmb{z}^{(1)}), \ldots, Y(\pmb{x}; \pmb{z}^{(n)}) \}$,
\[
\widehat{\mbox{Var}}[Y(\pmb{x}; \pmb{Z})] = \dfrac{1}{n-1}\sum_{i=1}^n [Y(\pmb{x}; \pmb{z}^{(i)}) - \bar{Y}]^2,
~~\mbox{with } \bar{Y} = \dfrac{1}{n}\sum_{i=1}^n Y(\pmb{x}; \pmb{z}^{(i)}).
\]

\textit{We are interested in quantifying the contribution of each random input $Z_{\ell}$ to output variation.} Here we explore the Shapley value (SV) based variance decomposition for IHSC risk analysis, which is motivated by game theory; see \cite{xie_submitted,song2016shapley,owen2014sobol}. 
The contribution from $Z_{\ell}$ can be measured by 
\begin{equation}\label{eq.SV}
s_{\ell} = \sum_{\mathcal{J}\subset \mathcal{L}/\{\ell\}} \dfrac{(L-|\mathcal{J}|-1)!|\mathcal{J}|!}{L!} \left[ c(\mathcal{J}\cup\{\ell\}) - c(\mathcal{J}) \right],
\end{equation}
where $\mathcal{L} = \{1, \ldots, L\}$ denotes the index set of random inputs, $|\cdot|$ indicates the set size, and cost function $c(\mathcal{J}) = \mbox{E}[\mbox{Var}[Y(\pmb{x}; \pmb{Z})| \pmb{Z}_{-\mathcal{J}}]]$ for any subset $\mathcal{J} \subset \mathcal{L}$, and $\pmb{Z}_{-\mathcal{J}}=\pmb{Z}_{\mathcal{L}/\mathcal{J}}$ denotes the remaining inputs. The SV in Equation~\eqref{eq.SV} has the property that $\mbox{Var}[Y(\pmb{x}; \pmb{Z})] = \sum_{\ell=1}^L s_{\ell}$. Therefore, it can be interpreted as the uncertainty/risk contribution from the $\ell$-th random input, which can guide end-to-end IHSC risk control.

Since we could not compute the analytical conditional variance and mean in cost function, we estimate $c(\mathcal{J})$ by simulation samples by using the procedure as follows. 
\vspace{-0.1in}
\begin{enumerate}
	\item For $k = 1, 2, \ldots, K$,
	\begin{itemize}
		\item[(1)] generate the $k$-th outer sample of random inputs $\pmb{z}_{-\mathcal{J}}^{(k)}$, 
		
		\item[(2)] For $i = 1, 2, \ldots, I$, generate the $i$-th inner sample of random inputs $\pmb{z}_{\mathcal{J}}^{(i)}$, and obtain simulation output realization $Y(\pmb{x}; \pmb{z}_{\mathcal{J}}^{(i)}, \pmb{z}_{-\mathcal{J}}^{(k)} )$
		
	\end{itemize}
	
	\item Then, the cost function $c(\mathcal{J}) = \mbox{E}[\mbox{Var}[Y(\pmb{x}; \pmb{Z})| \pmb{Z}_{-\mathcal{J}}]]$ can be estimated by,
	\begin{equation}\label{eq.c_hat}
	\widehat{c}(\mathcal{J}) = \dfrac{1}{K(I-1)}\sum_{k=1}^K \sum_{i=1}^I \left[ Y(\pmb{x}; \pmb{z}_{\mathcal{J}}^{(i)}, \pmb{z}_{-\mathcal{J}}^{(k)}) - \bar{Y}^{(k)} \right]^2,~~\mbox{with}~~
	\bar{Y}^{(k)} = \dfrac{1}{I}\sum_{i=1}^I Y(\pmb{x}; \pmb{z}_{\mathcal{J}}^{(i)}, \pmb{z}_{-\mathcal{J}}^{(k)}).
	\end{equation}
	
\end{enumerate}

\vspace{-0.1in}
\noindent By plugging  $\widehat{c}(\mathcal{J})$ by eq.~\eqref{eq.c_hat} into  eq.~\eqref{eq.SV}, we can obtain the estimation of SV. An efficient approximation algorithm (which uses the randomly selected subset instead of all possible index sets permutations) can be used to reduce the computational burden; see \cite{xie_submitted} and \cite{song2016shapley}.

\end{sloppypar}

\vspace{-0.1in}

\section{EMPIRICAL STUDY}
\label{sec:empiricalStudy}

In this section, we conduct the simulation experiments to study the performance of blockchain-based IHSC in terms of improving safety, efficiency, transparency, and reliability. 
We use the fixed number of resources at each step: (1) for tranplant and harvest, $n_f = 10$; (2) for pre-harvest, $n_l = 10$; (3) for drying, $n_d = 3$; (4) for extraction,winterization and PLC, $n_p = 2$. We set the number of lots to be $n=50$. The growth rate of cannabinoid is calculated as $C_8/56 = 10.2\%/56 = 0.18\%$ per day, where $C_8$ repsents the cannabinoid content 8 weeks (56 days) after flowering, and we set $\lambda = 50\%$ (see \href{https://oregoncbdseeds.com/resources/}{Oregon CBD}). For simplification, we assume the hard limit of waiting time for transplant $L_t$ and dry $L_d$ as 2 days. In addition, to avoid the product destruction required by the regulation, suppose each participant could tamper data with probability $p_2 = 30\%$. 
For the two-layer blockchain, let the verification and confirmation times follow the exponential distributions, i.e., $F_v \sim exp(\mu_v)$ and $F_{c} \sim \exp(\mu_{c})$ with means $\mu_v=0.1$ and $\mu_{c}=0.05$. There are 2 shards with $n_s=4$ local validators separately and two state regulators $n_r = 2$ in charge of confirmation at the root chain.

\vspace{-0.1in}

\subsection{Study the Performance of Blockchain-Enabled IHSC}

In this section, we have the simulation experiments to study the performance of IHSC. 
We set the simulation warmup and run-length to be $200$ and $K=500$, and set the number of replications $J=100$.

\noindent \underline{\textbf{IHSC Security and Safety:}} 
\textit{We first show the blockchain-enabled IHSC can improve the safety.} Here we compare the performance of IHSC with and without blockchain. Set the voting panel size $m=1$. For the critical record (i.e., harvest data,
final product quality data),
the validators will have onsite visit to ensure the data integrity.
To assess the IHSC safety, we record three false pass rates: (1) for pre-harvest test, the expected percentage of lots with more than 0.3\% THC that aren't destroyed, $q_{fp} = \underset{K\to\infty}{\lim}\mbox{E}[K_{fp}/K]$; (2) for the 15-day harvest regulation, the expected percentage of lots completed harvest violating the 15 days requirement, $q_{fh} = 
\underset{K\to\infty}{\lim}\mbox{E}[K_{fh}/K]$; and (3) the expected percentage of lots with more than 0.05\% THC and reaching to customers, $q_{fq} =  \underset{K\to\infty}{\lim}\mbox{E}[K_{fq}/K]$, where $K_{fp}$, $K_{fh}$, $K_{fq}$ are the counts of corresponding false approval. The simulation results in Table~\ref{IHWSC:SecuImpor} illustrate that blockchain can greatly improve the IHSC safety. 

\vspace{-0.05in}

\begin{table}[hbt!]
\small
\centering
\caption{Simulation results of security improvement.}
\label{IHWSC:SecuImpor}
\begin{tabular}{|c|c|c|}
\hline
{Security}               & {With Blockchain} & {Without Blockchain}  \\ \hline
{False Pass Pre-Harvest} & {0$\pm$0\%} & {2.03$\pm$1.23\%} \\ \hline
{False Pass Harvest}     & {0$\pm$0\%} & {2.98$\pm$2.28\%}  \\ \hline
{Fake Qualified}         & {0$\pm$0\%} & {0.68$\pm$0.69\%}  \\ \hline
\end{tabular}
\end{table}

\vspace{0.0in}
\noindent \underline{\textbf{Efficiency of Interoperability and Data Verification:}}
\textit{Two-layer blockchain can leverage local authorities and state regulatory inspection resources to improve the IHSC verification efficiency and facilitate the scale-up.} Suppose the amount of state regulatory official resources assigned for verification is fixed.
We compare the traditional single chain with our two-layer chain with 2 shards. For the single chain, the verification time for each record follows the distribution, $F_s \sim exp(\mu_s) $ with mean $\mu_s = 0.15$. 
In Table~\ref{IHWSC:CybEiffImpor}, we record the expected verification/confirmation time for each lot. 
The results show that two-layer blockchain can improve the efficiency. 
In addition, the expected number of dropped lots 
due to too long waiting time for dry, also decreases. 
There are considerable lots of biomass that need to wait for verification and cannot be dried in time, which causes the loss. Thus, two-layer blockchain can improve the efficiency, reduce the cycle time and eliminate the economy loss.
\vspace{-0.05in}

\begin{table}[hbt!]
\small
\centering
\caption{Simulation results of cyber efficiency improvement.}
\label{IHWSC:CybEiffImpor}
\begin{tabular}{|c|c|c|}
\hline
{Scalability}            & {Two-Layer Blockchain}   & {Classical Blockchain} \\   \hline
{Mean Verification Time} & {0.56$\pm$0.03}   & {0.85$\pm$0.1}    \\ \hline
{SD Verification Time}   & {0.26$\pm$0.17}   & {0.7$\pm$0.26}    \\ \hline
{Finished   \#}          & {275.3$\pm$27}    & {242.28$\pm$29.91}   \\ \hline
{Dry Drop   \#}          & {39.42$\pm$33.29} & {66.18$\pm$34.34}  \\ \hline
\end{tabular}
\end{table}

\noindent \underline{\textbf{IHSC Transparency and Process Control:}}
Built on the transparency of IHSC provided by blockchain, every participant can  dynamically schedule their resources with more efficiency. We implement the simulations under two situations: (1) the stabilizers don't know how many lots of biomass will come and fix the resource of dry machine $n_d = 3$; and (2) they can dynamically change the resource to avoid the dry-drop. 
The simulation results recorded in Table~\ref{IHWSC:PhyEiffImpor} illustrate that blockchain can facilitate the process control, improve the efficincy, and reduce the loss.

\vspace{-0.05in}

\begin{table}[hbt!]
\small
\centering
\caption{Simulation results of physical efficiency improvement.}
\label{IHWSC:PhyEiffImpor}
\begin{tabular}{|c|c|c|}
\hline
{Information   Share} & {Fixed Resource} & {Dynamic Resource} \\ \hline
{Finished   \#}       & {275.3$\pm$27}       & {306.11$\pm$23.9} \\ \hline
{Dry Drop   \#}       & {39.42$\pm$33.29}    & {3.04$\pm$3.79}  \\ \hline
\end{tabular}
\end{table}

\vspace{-0.2in}

\subsection{IHSC Process Risk Analysis}

Based on the study presented in Section~\ref{sec:uncertainty}, we conduct the IHSC process risk analysis to quantify the impact of each random input on the CBD and THC levels of final product. 
For the CBD oil percentage, the variation comes from $L = 7$ \underline{random inputs} $\pmb{Z} = \{ \epsilon,t^{\prime},\epsilon^{\prime},Q,W,Q_u,Q_v\}$, including (1)
growth intrinsic uncertainty during cultivation $\epsilon$; (2) the period  $t^{\prime}$ between taking the pre-harvest sample and finishing the harvest; (3) growth intrinsic uncertainty from pre-harvest to harvest $\epsilon^{\prime}$; (4) extraction rate $Q$; (5) winterization rate $W$; (6) CBD removed rate during PLC $Q_u$; (7) THC removed rate during PLC $Q_v$.  
Particularly, the distribution and variability of $t^{\prime}$ is estimated by using the simulation.
The number of random subset permutations, $m=3000$, is used to estimate the SV.
In addition, the variation of THC level caused by $L = 6$ random inputs $\pmb{Z} = \{\epsilon,t^{\prime},\epsilon^{\prime},Q,W,Q_v\}$. Here, we consider all possible permutations $m = L! = 720$. We set the number of outer samples $K = 10$, the number of inner samples $I = 100$ and macro-replications $J = 10$ . The simulation results of relative contribution from each input factor are calculated by applying Equations~(\ref{eq.SV}) and (\ref{eq.c_hat}). 
The results from $J$ macro-replications, $RC_{\ell} = \frac{1}{J}\sum_{j=1}^J\frac{s_{\ell}^{(j)}}{\widehat{\mbox{Var}}[Y(\pmb{x}; \pmb{Z})]}$, are represented in Table~\ref{IHWSC:IHSCSV}. 
Clearly, the most part of lot-to-lot variation of both CBD and THC comes from the cultivation step.

\vspace{-0.05in}

\begin{table}[hbt!]
\small
\centering
\caption{The estimated relative contribution of each random input factor}
\label{IHWSC:IHSCSV}
\begin{tabular}{|c|c|c|c|c|c|c|c|}
\hline
{Input} & {$\epsilon$} & {$t^{\prime}$} & {$\epsilon^{\prime}$} & {$Q$} & {$W$} & {$Q_u$} & {$Q_v$} \\ \hline
{CBD \%}   & {69.8$\pm$5.3} & {3.0$\pm$1.8} & {8.0$\pm$3.0} & {4.9$\pm$1.5} & {4.2$\pm$1.4} & {4.4$\pm$2.9} & {5.4$\pm$2.8} \\ \hline
{THC \%}   & {60.9$\pm$3.2} & {2.4$\pm$1.8} & {14.8$\pm$3.6} & {6.9$\pm$2.6} & {1.5$\pm$2.2} & {N.A.} & {15.0$\pm$2.0} \\ \hline
\end{tabular}
\end{table}

\vspace{-0.1in}
\section{CONCLUSION}
\label{sec:conclusion}

In this paper, we propose the blockchain-enabled IHSC and introduce the simulation-based digital twin to: (1) improve the IHSC transparency, safety, security, and efficiency; (2) conduct the process risk analysis and guide the decisions. The empirical results indicate that the proposed platform has promising performance. 
\textit{For the future research, based on the real data collected in our project, we will focus on: (1) providing deep, interpretable, and data-efficient IHSC process learning (e.g., how the various factors, including the seed selection, soil and weather, impact on the IH production); (2) proposing the knowledge-based participants' reputation learning; and (3) improving and validating the digital twin so that it can provide high fidelity of real-world blockchain-enabled IHSC risk behaviors; and (4) accelerating the end-to-end IHSC automation.}

\section*{ACKNOWLEDGMENTS}
The authors acknowledge the invaluable feedback from Oregon Department of Agriculture and USDA.


\footnotesize

\bibliographystyle{unsrt}

\bibliography{references}

\end{document}